\documentstyle[graphics]{article}
\begin{document}

\def\la{\mathrel{\mathpalette\fun <}}
\def\ga{\mathrel{\mathpalette\fun >}}
\def\fun#1#2{\lower3.6pt\vbox{\baselineskip0pt\lineskip.9pt
\ialign{$\mathsurround=0pt#1\hfil##\hfil$\crcr#2\crcr\sim\crcr}}}  
\def\lrang#1{\left\langle#1\right\rangle}

\begin{center}
{\Large \bfseries Jet quenching in heavy ion collisions at LHC~\footnote{Talk 
given at XXXII International Symposium on Multiparticle Dynamics, Alushta,
Crimea, September 7-13, 2002.}} 

\vskip 4mm

I.P. Lokhtin$^{\dag}$

\vskip 4mm

{\small{\it M.V.Lomonosov Moscow State University, D.V.Skobeltsyn Institute of Nuclear 
Physics}\\ 
$\dag$ {\it E-mail: igor@lav01.sinp.msu.ru}}
\end{center}

\vskip 4mm

\begin{center}
\begin{minipage}{150mm}
\centerline{\bf Abstract}

\smallskip 

We discuss the potential information about highly excited QCD-matter provided by 
medium-induced partonic energy loss, known as ``jet quenching''. In particular, 
with its large acceptance hadronic and electromagnetic calorimetry, the Compact Muon 
Solenoid detector at LHC collider is a promising device to study these effects. We present 
physics simulations of observables such as the jet distribution with impact parameter, the 
azimuthal anisotropy of jet quenching, and the effects of $b$-quark energy loss on the high-mass 
dimuon continuum and secondary $J/\psi$ production. \\

\end{minipage}
\end{center}

\section{Hard probes at LHC with CMS detector}

High-$p_T$ QCD phenomena are perspective tools to study the properties of the 
hot matter created in ultrarelativistic nuclear interactions~\cite{hard_probes}. Hard probes such 
as quarkonia, hard jets and heavy quarks are 
not thermalized but carry information about the earliest stages of a system subject to 
numerous final-state effects in heavy ion collisions at SPS and RHIC.   
At the LHC, a new regime is reached where hard and semi-hard QCD 
multi-particle production can already dominate over underlying soft events.  

The following signals of jet quenching due to medium-induced parton energy 
loss~\cite{Baier:2000}  can be identified as being observable in heavy ion collisions at  
LHC~\cite{hard_probes}. \\
1) The suppression of high-$p_T$ jet pairs as compared to what is expected from 
independent nucleon-nucleon interactions pattern~\cite{Gyulassy:1990}. It includes 
modification of jet impact parameter dependence~\cite{Lokhtin:2000} and 
azimuthal anisotropy of jet quenching in non-central collisions~\cite{Lokhtin:2001}.  \\ 
2) The $p_T$-imbalance in 
$\gamma+$jet~\cite{Wang:1996} and $Z(\rightarrow \mu^+ \mu^-)+$jet~\cite{Kartvelishvili:1996} 
channels. \\   
3) The suppression of high-$p_T$ hadron and photon (leading particle) yields due to a 
modification of the jet fragmentation function~\cite{Gyulassy:1992}. \\ 
4) The modification of the high mass dimuon spectra from semileptonic $B$ and $D$ meson 
decays and secondary charmonium production due to heavy quark energy 
loss~\cite{Lin:1999,Lokhtin:2001hq}. 

All above signals can be studied in CMS Heavy Ion Programme~\cite{Baur:2000}. 
The Compact Muon Solenoid (CMS) is a general purpose detector designed primary to 
search for the Higgs boson in proton-proton collisions at LHC~\cite{CMS:1994}. 
Accordingly, the detector is optimized for 
accurate measurements of the characteristics of high-energy leptons, photons 
and hadronic jets in a large acceptance, providing unique
capabilities for ``hard probes'' for both $pp$ and $AA$ collisions.
The central element of CMS is the magnet, a $13$ m long and 
$6$ m diameter solenoid with an internal radius $\approx 3$ m, which will provide 
a strong $4~T$ uniform magnetic field. The $4\pi$ detector consists of a $6$ m 
long and $1.3$ m radius central tracker, electromagnetic (ECAL) and hadronic
(HCAL) calorimeters inside the magnet and muon stations outside. 
The tracker and muon chambers covers the pseudorapidity $|\eta|<2.4$, while the ECAL and 
HCAL calorimeters reach $|\eta|=3$. A pair of quartz-fiber very forward (HF) calorimeters, 
located $\pm 11$ m from the interaction point, cover the region $3<|\eta|<5$ 
and complement the energy measurement. The tracker is composed of pixel layers 
and silicon strip counters. The CMS muon stations consist of drift tubes in the barrel
region and cathode strip chambers in the end-cap regions. The electromagnetic 
calorimeter is made of almost 
$83000$ scintillating ${\rm PbWO_4}$ crystals, and the hadronic calorimeter consists of 
scintillator inserted between copper absorber plates. The main characteristics of
calorimeters are presented in Table 1. 
 
Table 2 presents the event rates for various channels including hard
jets in one month of Pb$-$Pb beams assuming two weeks of data taking and  
luminosity $L = 5 \times 10^{26}~$cm$^{-2}$s$^{-1}$.  The production cross sections in
minimum bias $AA$ collisions were obtained from those in $pp$ interactions 
at the same energy, $\sqrt{s} = 5.5$ TeV, using simple parameterization 
$\sigma^h_{AA}=A^2 \sigma^h_{pp}$. The $pp$ cross sections were evaluated
using the PYTHIA\_$6.1$~\cite{pythia} with the CTEQ5L parton 
distribution functions. The estimated for the CMS acceptance statistics 
will be large enough to carefully study the dijet rate as a function of
impact parameter as well as the $\varphi$ and $\eta$ distributions of jets. 
The statistics for the $\gamma$+jet channel are satisfactory for studying 
the $E_T$-imbalance of the process, but the large background from 
jet$+$jet$(\rightarrow \pi^0)$ is still under investigation. The 
statistics for $Z (\rightarrow \mu^+ \mu^-)$+jet channel are rather low, 
but the background is less than $10\%$ in this case.
Note that jet  reconstruction in heavy ion 
collisions at CMS with efficiency and purity close to $100\%$ is possible
starting $E_T^{jet} \sim 100$ GeV with energy resolution worser by factor $\sim 2$
as compared to $pp$ case~\cite{Baur:2000}. 

\begin{table}
\begin{center}
\label{tab1} 
\caption{\small Energy resolution $\sigma /E = a/\sqrt{E} \bigoplus b$ and granularity of CMS 
calorimeters in barrel (HB, EB), endcap (HE, EE) and very forward (HF) regions. The energy 
resolution is shown for the total energy of $e$ and $\gamma$ (EB, EE) and transverse
energy of hadronic jets (HB, HE, HF).} 

\medskip 

\begin{tabular}{|l|c|c|c|c|c|} \hline  
Rapidity & \multicolumn{2} {c|} {$0<\mid\eta\mid<1.5$} & \multicolumn{2} {c|} 
 {$1.5<\mid\eta\mid<3.0$} 
  & {$3.0<\mid\eta\mid<5.0$} \\ \hline 
Subdetector & HB & EB & HE & EE & HF \\ \hline 
$a$ & 1.16 & 0.027 & 0.91 & 0.057 & 0.77 \\
$b$ & 0.05 & 0.0055 & 0.05 & 0.0055 & 0.05 \\
\hline
granularity &  &  &  &  0.0174$\times$0.0174&
\\
$\Delta\eta \times \Delta\varphi$ & 0.087$\times$0.087& 0.0174$\times$0.0174 & 
0.087$\times$0.087 &  to 0.05$\times$0.05 & 0.175$\times$0.175 \\\hline 
\end{tabular}
\end{center}
\end{table}

\begin{table}
\begin{center}
\vskip -5 mm  
\label{tab2} 
\caption{\small Expected rates for various channels including hard jets in one month of 
Pb$-$Pb beams.} 

\medskip 

\begin{tabular}{|l|c|c|} \hline  
Channel & Barrel & Barrel+Endcap  
\\ \hline 
jet+jet, $E_T^{\rm jet}>100$ GeV & 2.1$\times$10$^6$ & 4.3$\times$10$^6$ 
 \\ \hline   
$\gamma$+jet, $E_T^{{\rm jet},\gamma}>100$ GeV & 1.6$\times$10$^3$ & 
3.0$\times$10$^3$ 
\\ \hline 
$Z (\rightarrow \mu^+ \mu^-)$+jet, $E_T^{\rm jet},P_T^{Z}>100$ GeV & 30 & 45 \\    
\hline 
\end{tabular}
\end{center}
\end{table}

\section{Jets versus impact parameter and azimuthal angle}  

\begin{figure}[hbtp]
\begin{minipage}[t]{75mm}
\resizebox{74mm}{74mm}  
{\includegraphics{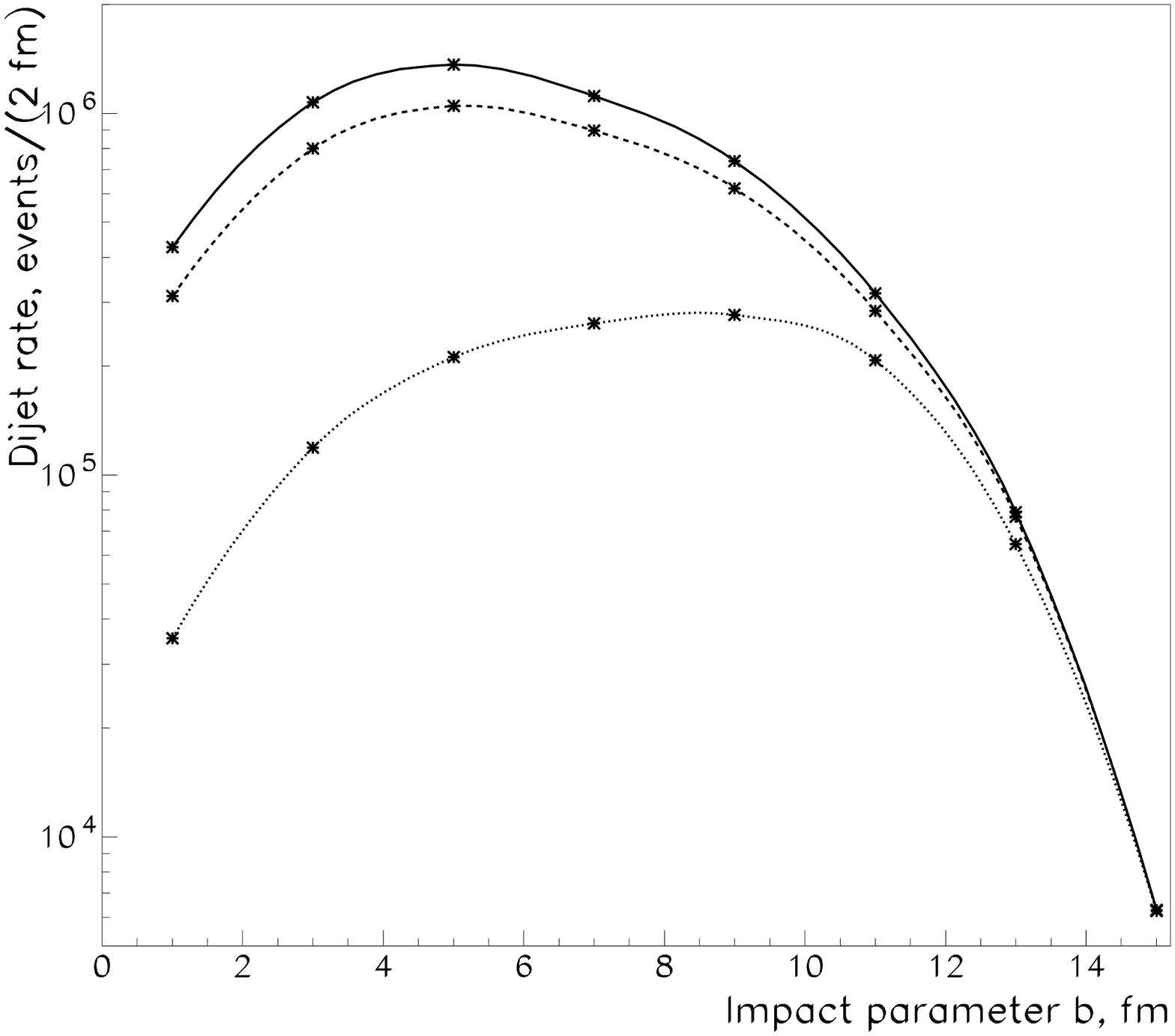}} 
\caption{\small The dijet rates vs. 
impact parameter $b$: without energy loss (solid curve), with 
collisional loss (dashed), with collisional and radiative loss (dotted).} 
\label{ng:fig4} 
\end{minipage}
\hskip 3 mm 
\begin{minipage}[t]{75mm} 
\resizebox{74mm}{74mm} 
{\includegraphics{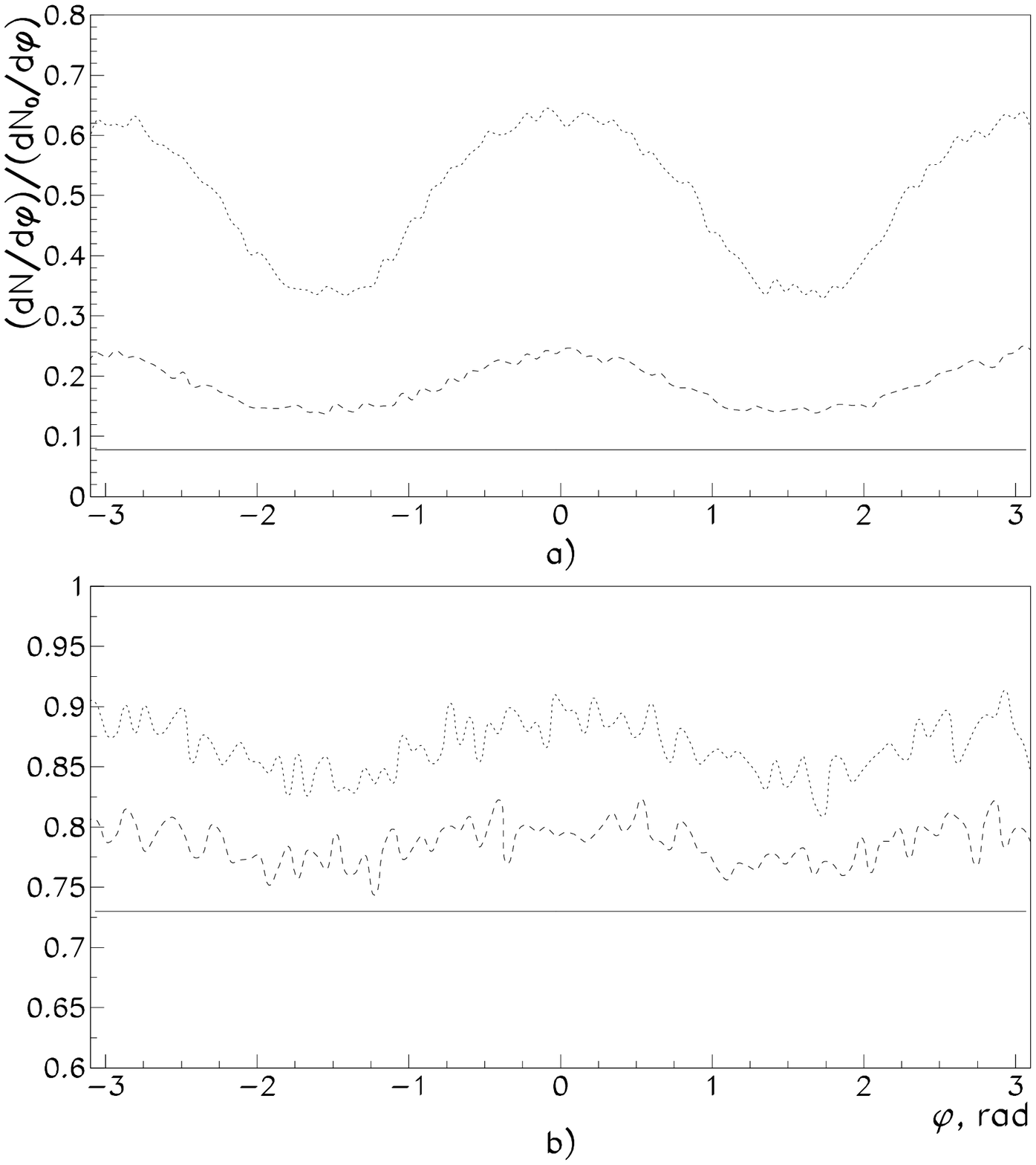}} 
\caption{\small The distribution of jets over azimuthal angle: collisional and radiative loss (a), 
collisional loss only (b). From bottom to top: $b = 0$, $6$ and $10$ fm.} 
\label{ng:fig5}
\end{minipage}
\end{figure}

We have analyzed the impact parameter dependence of jet rates in Pb$-$Pb collisions at LHC 
(see Ref.~\cite{Lokhtin:2000} for model details).  
Figure~\ref{ng:fig4} shows dijet rates in different impact parameter bins  
($E_T^{jet} > 100$ GeV, $|\eta^{jet}| < 2.5$): $(i)$ without 
energy loss, $(ii)$ with collisional loss only, $(iii)$ with collisional and radiative loss 
from BDMS model~\cite{Baier:1998n}. The bin width $2$ fm in Fig.~\ref{ng:fig4}  
corresponds to estimated accuracy of impact parameter determination using correlation between 
$b$ and energy flow deposited in very forward CMS HF calorimeter~\cite{Damgov:2001}. The 
maximum and mean values of $dN^{dijet}/db$ distribution get shifted towards the larger $b$ 
because jet quenching is stronger in central collisions than in peripheral one's,. Since the 
coherent LPM radiation induces a strong dependence of the radiative 
energy loss of a jet on the angular cone size~\cite{Baier:1998,Lokhtin:1998}, 
the result for jets with non-zero cone size $\theta_0$ is expected to 
be between $(iii)$ ($\theta_0 \rightarrow 0$) and $(ii)$ cases. Thus the  
observation of a dramatic change in the $b$-dependence of jet rates 
as compared to what is expected from independent nucleon-nucleon 
interactions pattern, would indicate the existence of medium-induced parton 
rescattering. 

Other interesting feature of jet production in semi-central $AA$ collisions can be azimuthal
anisotropy of high-$p_T$ hadrons~\cite{Wang:2000,Gyulassy:2000} or jets~\cite{Lokhtin:2001} 
due to parton energy loss in azimuthally non-symmetric 
volume of quark-gluon plasma. Figure~\ref{ng:fig5}  shows the distribution of 
jets over azimuthal angle $\varphi$ with collisional and radiative 
loss (a) and collisional loss only (b) for $b = 0$, $6$ and $10$ fm. The same 
conditions as for Fig.~\ref{ng:fig4} are fulfilled.  
The plot is normalized on the distributions of jets over
$\varphi$ in Pb$-$Pb collisions without energy loss. 
The azimuthal anisotropy gets stronger as going from central to semi-central
collisions, but the absolute suppression factor reduces with increasing $b$. 
The methodical advantage of azimuthal jet observables is that one needs to 
reconstruct only azimuthal position of jet without measuring total jet energy. 
The summarized in papers~\cite{Voloshin:1996} methods for 
determination of the event plane are applicable for studying anisotropic flow 
in current heavy ion experiments at SPS and RHIC, and might be also 
used at LHC~\cite{Lokhtin:2001}. It was also suggested recently in~\cite{Lokhtin:2002} the 
method for measurement of jet azimuthal anisotropy without reconstruction 
of the event plane. This technique is based on the calculation of correlations between 
the azimuthal position of jet axis and the angles of particles (not incorporated in the 
jet) using as weights the particle momenta or energy deposition in the calorimetric sectors.

\begin{figure}[hbtp]
\begin{center} 
\resizebox{120mm}{120mm} 
{\includegraphics{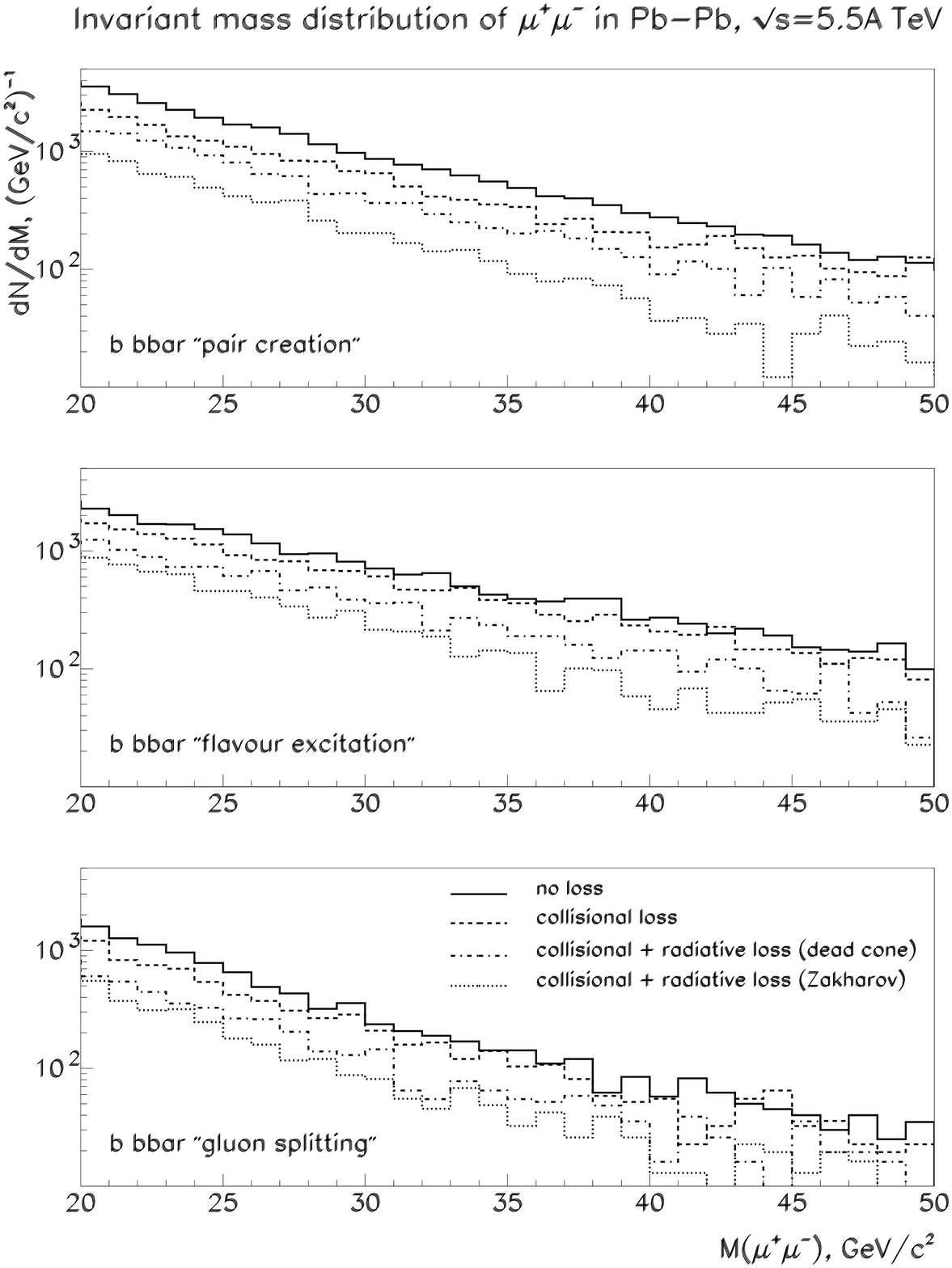}} 
\caption{\small Invariant mass distribution of $\mu^+\mu^-$ pairs  from $B \overline{B}$ decays 
for several models of energy loss. The muon requirements are $p_T^{\mu} > 5$
GeV/c and $|\eta^{\mu}|<5$.} 
\label{fig:hq2}
\end{center} 
\end{figure}

\section{Heavy quarks: $B{\bar B} \rightarrow \mu^+\mu^-$ and $B \rightarrow 
J/\psi $ modes}  

The open heavy flavour production is important due to opportunity to 
study the massive colour charge in quark-gluon environment, while jets 
probe the behaviour of massless partons in QCD-matter. The charm and bottom
production cross section are much larger at LHC as compared with RHIC. Thus the 
systematical studies of heavy flavour (especially $B$) physics can be performed 
in CMS. It was found in the high invariant mass dimuon range 
$10$ GeV/c$^2 \leq M \leq 70$ GeV/c$^2$ the dominant contribution comes from $B \bar B$ 
fragmentation, and its extraction from uncorrelated and correlated background
is possible~\cite{Baur:2000,Lokhtin:2001hq}. In-medium gluon radiation and collisional energy 
loss of heavy quarks can result in suppression and modification of high mass 
dilepton~\cite{Lin:1999,Lokhtin:2001hq} and secondary $B
\rightarrow J/\psi$~\cite{Lokhtin:2001hq} spectra. In addition, finite quark mass effects can 
lead to the suppression of medium-induced radiation of heavy quarks as compared with
massless partons, enhancing e.g. the $B/\pi$ ratio~\cite{Dokshitzer:2001}. 

The important predicted feature of heavy quark production at LHC 
is that the contribution of gluon splittings in initial- or final-state shower 
evolution to heavy flavour yield can be important~\cite{Norrbin:2000}, while at 
present accelerator energies the most of heavy quarks being produced due to direct hard 
scatterings (``pair creation''). These are so-called ``flavour excitation'' (one heavy quark is 
produced in vertex of hard process and another quark is created from initial state parton shower) 
and ``gluon splitting'' (both heavy quarks are produced from final state parton shower). It was 
shown in~\cite{Lokhtin:2001n} that such ``showering'' $b \bar{b}$ pairs are an 
important contribution to the dilepton continuum at high masses and, in fact, are of 
the same order as direct $q\bar{q}\rightarrow Q\bar{Q}X$ and $gg\rightarrow Q\bar{Q}X$ 
pair creation at the LHC for dimuons with masses greater than $10$ GeV/c$^2$. Moreover, 
most of secondary $J/\psi$'s produced in $B$ decays originate from the $gg \rightarrow
g(g^{*}\rightarrow Q\bar{Q})$ diagram~\cite{Lokhtin:2001n}.  

Figure~\ref{fig:hq2} shows the dimuon invariant mass spectra 
at CMS acceptance for several models: without loss;
with rescattering and collisional loss; with rescattering, collisional and radiative
loss calculated as incoherent limit of Zakharov's formula~\cite{Zakharov}, and also
from ``dead-cone'' model~\cite{Dokshitzer:2001}. The suppression factor $\sim 2$
-$4$ can be clearly observed over ``nuclear shadowing'' effect~\cite{EKS}, $\sim 15 \%$. 
Estimated suppression by a factor 
of $\sim 1.5$-$2$ for secondary $J/\psi$ is less than for $B \overline{B}$ decays and
is comparable with the shadowing corrections ($\sim 30 \%$).  

\section{Conclusion}

To summarize, with its large acceptance, almost hermetic hadronic and electromagnetic 
calorimetry of fine granularity and good energy and spatial jet resolution, and excellent muon
capabilities, the Compact Muon Solenoid (CMS) detector is a promising device to study jet 
quenching at LHC in various channels. Jet  reconstruction in heavy ion 
collisions at CMS is possible
starting $E_T^{jet}=100$ GeV, and estimated statistics will be large enough to carefully study the 
jet rate as a function of impact parameter, as well as azimuthal and rapidity distributions of
jets. Impact parameter and  event plane determination are possible using energy flow 
measurements. In-medium energy loss of heavy quarks also can be observable in 
$B{\bar B} \rightarrow \mu^+\mu^-$ and $B \rightarrow J/\psi \rightarrow \mu^+\mu^- $ modes.
However, in order to determine the baseline rates precisely, measurements in $pp$ or $dd$ 
collisions at the same or similar energies per nucleon as in the heavy ion runs are strongly 
desirable. 

{\it Acknowledgements.}  
I would like to thank the organizers of the Symposium for the warm welcome and stimulating 
atmosphere. Discussions with M.~Bedjidian, D.~Denegri,  Yu.L.~Dokhitzer, D.~Kharzeev, 
O.L.~Kodolova, A.N.~Nikitenko, S.V.~Petrushanko, L.I.~Sarycheva, S.V.~Shmatov, A.M.~Snigirev,  
V.V.~Uzhinskii, I.N.~Vardanian, S.A.~Voloshin, R.~Vogt, U.~Wiedemann, P.I.~Zarubin and
G.M.~Zinoviev are gratefully acknowledged.

\end{document}